\documentclass[twocolumn]{aastex62}
\pdfoutput=1 %for arXiv submission
\usepackage{amsmath,amstext}
\usepackage[T1]{fontenc}
\usepackage{apjfonts} 
\usepackage[figure,figure*]{hypcap}

\shorttitle{The fate of the merger remnant  in GW170817  and its imprint on jet structure } 
\shortauthors{Murguia-Berthier et al.}

\begin{document}
\title{The fate of the merger remnant  in GW170817  and its imprint on the jet structure}
\author{Ariadna~Murguia-Berthier}
\affiliation{Department of Astronomy and Astrophysics, University of California, Santa Cruz, CA 95064, USA}
\affiliation{DARK, Niels Bohr Institute, University of Copenhagen, Blegdamsvej 17, 2100 Copenhagen, Denmark}
\author{Enrico~Ramirez-Ruiz}
\affiliation{Department of Astronomy and Astrophysics, University of California, Santa Cruz, CA 95064, USA} 
\affiliation{DARK, Niels Bohr Institute, University of Copenhagen, Blegdamsvej 17, 2100 Copenhagen, Denmark}
\author{Fabio~De Colle}
\affiliation{Instituto de Ciencias Nucleares, Universidad Nacional Aut\'onoma de M\'exico, A. P. 70-543, 04510 CDMX, M\'exico}
\author{Agnieszka~Janiuk}
\affiliation{Centrum Fizyki Teoretycznej PAN
Al. Lotnik\'ow 32/46, 02-668 Warsaw, Poland}
\author{Stephan~Rosswog}
\affiliation{Astronomy and Oskar Klein Centre, Stockholm University, AlbaNova, SE-10691, Stockholm, Sweden }
\author{William~H.~Lee}
\affiliation{Instituto de Astronom\'ia, Universidad Nacional Aut\'onoma de M\'exico, A. P. 70-264, 04510 CDMX, M\'exico}

\begin{abstract}
The first neutron star binary merger detected in gravitational waves, GW170817 and the subsequent detection of its emission across the electromagnetic spectrum showed that these systems are viable progenitors of short $\gamma$-ray bursts (sGRB). The afterglow signal of GW170817 has been found to be consistent with a structured GRB jet seen off-axis, requiring significant amounts of relativistic material at large angles. This trait can be attributed to the interaction of the relativistic jet with the external wind medium. Here we perform numerical simulations of relativistic jets interacting with realistic wind environments in order to explore how the properties of the wind and central engine affect the structure of successful jets. We find that the angular energy distribution of the jet depends primarily on the ratio between the lifetime of the jet and the time it takes the merger remnant to collapse. We  make use of these  simulations  to  constrain the  time it took for  the merger remnant in  GW170817 to collapse into a black hole based on the angular structure of the jet as inferred from afterglow observations. 
 We conclude that the lifetime of the merger   remnant in GW170817 was  $\approx 1-1.7$s, which, after collapse, triggered the formation of the jet.
\end{abstract}
\keywords{gamma-ray burst: GRB170817, jets- hydrodynamics}

\section{Introduction }
The gravitational wave event GW170817 \citep{2017PhRvL.119p1101A}, that was accompanied by the detection of  emission  across the electromagnetic spectrum \citep{2017Sci...358.1556C,2017ApJ...848L..12A}, demonstrated that neutron star binary mergers are the sources of short $\gamma-$ray bursts \citep[sGRBs;][]{Eichler1989,1992ApJ...395L..83N}, or at least a subset of them. 

From the unusually faint nature of GRB 170817A it was initially argued that this event belonged to a  class of intrinsically sub-energetic sGRBs \citep{2018ApJ...856L..18M}, but as more is being learned about GRB 170817A, the more it appears like the afterglow emission  is  instead consistent with the neutron star merger having triggered a typical, powerful sGRB seen at an angle of about a few times the opening angle of the central jet \citep{2017MNRAS.471.1652L, 2017ApJ...848L..34M,2018ApJ...866....3D,2018MNRAS.481.2581L,2018Natur.561..355M,2018ApJ...869...55W,2018ApJ...868L..11M,2019ApJ...870L..15L}. 

A question that has remained largely unanswered so far is what determined the atypical structure of the  jet  in GRB 170817A, which is required  not to have sharp edges but wings of lower kinetic energy and Lorentz factors that extend to large angles \citep{2017MNRAS.471.1652L,2018ApJ...866....3D,2018MNRAS.475.2971B,2018MNRAS.481.1597G,2018MNRAS.478..733L,2018ApJ...863...58X,2018MNRAS.473L.121K,2018arXiv180806617V,2019ApJ...883...15G,2019ApJ...870L..15L,2019arXiv190408425L,2020MNRAS.493.3521B,2020ApJ...898...59L,2020ApJ...896..166R,2020arXiv200501754N}. This in turn might  be attributed to the interaction of the jet with a dense surrounding gas distribution, quite possibly the wind emanating from the merger remnant. 

A trait of the jet-wind interaction  is that there can be an exchange of linear momentum with the neighboring wind, even if there is minimal exchange of baryons \citep{Lee07}. As a result, the  jet itself is then likely to develop a velocity profile so that different portions  move with different Lorentz factors \citep{2005A&A...436..273A,2002MNRAS.336L...7R}.  This implies that an observer could infer a value for the Lorentz factor that depends upon the inclination of the line of sight to the jet axis.  In this case, the early afterglow emission of  GRB 170817A
would  be naturally produced by the deceleration  of the lower Lorentz factor material  moving along our line of sight \citep{2003MNRAS.345.1077R, 2018MNRAS.478..733L}.  

We already know that winds are a prominent feature of neutron star merger remnants  and are  accompanied by prodigious mass loss \citep{2002MNRAS.336L...7R, 2003MNRAS.343L..36R,Lee07,2009ApJ...699L..93L, 2014MNRAS.443.3134P,2020arXiv200804333N}. This is not on the face of it the most advantageous  environment to produce  a  baryon-starved  jet. Studies of sGRB jets \citep{Eichler1989,1992ApJ...395L..83N,Lee07}  are thus increasingly invoking  the collapse of the  remnant  to a black hole \citep[e.g.][]{2005ApJ...630L.165L}, which  neatly avoids the problem of catastrophic mass pollution  close to the jet creation region, and allows the terminal jet Lorentz factors to be large.  There are alternatives to the collapse of the remnant to a BH in which a long-lived remnant can form a jet \citep{2020arXiv200306043M}, although the aforementioned risks of baryon population could severely limit the jet's Lorentz factor. These issues motivate our study of the interaction between the jet and the pre-collapse wind  through which the jet is expected to propagate. This jet-outflow interaction could turn out to be key for interpreting the observed properties of GRB 170817A. 

The pre-burst gas distribution depends on how the binary neutron star merger loses mass.  As the binary coalesces, various mechanisms can transport angular momentum and dissipate energy in the newly formed remnant  \citep{2008PhRvD..78h4033B}, giving rise to significant mass-loss \citep{Lee07}. One  important transport mechanism is the neutrino-driven wind (NDW) \citep{2003MNRAS.343L..36R,dessart09,2014MNRAS.443.3134P}, as copious amounts of neutrinos are created in the merger remnant. The high flux of neutrinos will interact with matter in and around the merger remnant, driving a baryon loaded wind. Another important mechanism is energy and angular momentum   transport mediated by magnetic fields. Even an initial  weak field can destabilize the merger remnant via the so called magneto-rotational instability \citep[e.g.][]{2019PhRvD..99h4032R}.  Magnetic instabilities inside the merger remnant can thus drive a powerful baryon-loaded wind  \citep{Siegel14, 2017PhRvD..95f3016C,2020MNRAS.495L..66C}.

This merger remnant is assumed to be unstable and its subsequent collapse to a black hole is assumed to trigger a relativistic jet \citep{2011ApJ...732L...6R,2012MNRAS.423.3083M, 2014MNRAS.441.3177M, 2015MNRAS.447...49S, 2016ApJ...824L...6R, 2018arXiv180806617V,2018ApJ...859...28Q,2019MNRAS.484L..98K,2019PhRvD..99h4032R}. 
This is thought to be  the case for GRB 170817, since a long-lived remnant is difficult to reconcile with observations \citep[e.g.][]{2020MNRAS.495L..66C,2019PhRvD.100b3005C}. In this case, the jet will then unavoidably interact with the pre-collapse winds\footnote{As the jet continues to plough ahead of the wind, it sweeps up an increasing amount of surrounding gas, made up of interstellar medium  or  material which was previously ejected by the progenitor system \citep[such as, e.g.,  a pulsar-wind;][]{2019ApJ...883L...6R}. This sets a characteristic deceleration length at which the the afterglow emission becomes relevant.}. 

The interaction of the wind has a decisive effect in shaping the jet \citep{2014ApJ...784L..28N,ari14,2015ApJ...813...64D,ari17,2018ApJ...866....3D,2018PhRvL.120x1103L,2019ApJ...877L..40G,2020arXiv200710690H,2020arXiv200602466G,2020arXiv200711590G,2020MNRAS.491.3192H}. Additionally, there are instances in which the wind can be dense enough to become a death trap for the jet, choking the outflow and rendering  sGRB production unsuccessful. Therefore it is important to understand how the properties of the wind and central engine affect the structure of the jet.   The wind is driven by the merger remnant  in this case while the jet  is triggered by the newly formed BH and accretion disk. A key property of the binary neutron star merger is the time it takes for the remnant to collapse into a black hole, which is directly linked to the lifetime of the wind.   Thus, by understanding how the interaction with the wind can alter the jet structure, one can constrain the lifetime of this remnant.

In this paper we perform special relativistic simulations to study how the jet structure is affected by the neutrino driven wind and the magnetized disk outflow. We also explore how the  time delay between collapse and jet triggering affects the structure of the jet. We use these simulations to set a limit on the delay time of GW170817 based on the angular structure of the jet that is inferred from  afterglow observations.

\section{Jet and wind interaction}

\subsection{Numerical method and setup}
Our simulations follow the setup described in \citet{ari14,ari17}. They are performed in 2D axisymmetric coordinates using \textit{Mezcal}, an adaptive mesh refinement code that solves the equations of special relativistic hydrodynamics.  A description of the code and a number of benchmark
tests can be found in \citet{2012ApJ...746..122D,2012ApJ...751...57D}. 

The setup begins with the injection of a wind, lasting for a time $t_{\rm w }$. This $t_{\rm w}$ is directly related to  the time it takes the merger remnant  to collapse to a black hole. After that time, the density of the wind is assumed to decrease as $t^{-5/3}$, and a jet is introduced. Motivated by GRMHD simulations \citep{2012MNRAS.423.3083M,2015MNRAS.447...49S,  2016ApJ...824L...6R,2019PhRvD..99h4032R}, the jet is assumed to initially have a top-hat structure characterized by a half-opening angle $\theta_0$, a luminosity $L_j$ and a Lorentz factor of $\Gamma=10$ that are constant with angle\footnote{ Numerical simulations of jets launched from the central engine with a luminosity and Lorentz factor varying with the polar angle $\theta$ are discussed in \citep{2020arXiv201106729U}.}. The central engine powers the jet for a time $t_{\rm j}$. Table~\ref{table:data} presents  the simulation parameters of all the calculations we performed in this study.

\begin{table*}[h!]
\centering
\resizebox{0.9\textwidth}{!}{
\begin{tabular}{||c c c c c c c||} 
 \hline
 Type of wind & $t_w$ (s)& $t_j (s)$  & $L_{\rm j}/10^{50}$ (erg/s) & $\dot{M}_{\rm w}/10^{-3}$ ($M_\odot$/s) &$\theta_0$ ($^\circ$)   & Successful sGRB?  \\ [0.5ex] 
 \hline\hline
 SW & 0.5 & 0.5  & 1 & 1 & 10   & Yes \\ 
 SW & 1 & 0.5  & 1 & 1 & 10     & Mild \\ 
 SW & 0.5 & 1  & 1 & 1 & 10   & Yes \\ 
 SW & 1 & 1  & 1 & 1 & 10    & Yes \\ 
 SW & 0.5 & 0.5  & 1 & 10 & 10   & No \\
 SW & 0.5 & 1  & 1 & 10 & 10 & No \\ 
 SW & 1 & 0.5  & 1 & 10 & 10  & No \\ 
 SW & 0.5 & 0.5  & 10 & 10 & 10   & Yes \\
 SW & 0.3 & 0.5  & 10 & 10 & 10   & Yes \\ 
 SW & 1 & 0.5  & 10 & 10 & 10   & Mild \\ 
 SW & 0.5 & 1  & 10 & 10 & 10    & Yes \\ 
 SW & 0.3 & 1  & 1 & 5 & 14    & Yes \\ 
 SW & 0.7 & 1  & 1 & 1 & 10    & Yes \\ 
 NDW & 0.5 & 0.5  & 1 & 1 & 10   & Yes \\ 
 NDW & 0.5 & 1  & 1 & 1 & 10  & Yes \\
 NDW & 1 & 0.5  & 1 & 1 & 10   & Mild \\
 NDW & 1 & 1  & 1 & 1 & 10   & Yes \\
 NDW & 0.5 & 0.5  & 1 & 3 & 10     & Yes\\
 NDW & 0.5 & 0.5  & 5 & 3 & 10    & Yes \\
 NDW & 0.5 & 0.5  & 1 & 10 & 10   & No \\
 NDW & 0.5 & 1  & 1 & 10 & 10    & No \\
 NDW & 1 & 0.5  & 1 & 10 & 10   & No \\
 NDW & 1 & 1  & 1 & 10 & 10     & No \\
 NDW & 0.5 & 0.5  & 10 & 10 & 10 & Yes \\
 NDW & 0.5 & 1  & 10 & 10 & 10 & Yes \\
 NDW & 1 & 0.5  & 10 & 10 & 10   & Mild \\
 NDW & 1 & 1  & 10 & 10 & 10 & Yes \\
 NDW & 0.3 & 1  & 1 & 1 & 10  & Yes \\
 NDW & 0.3 & 1  & 10 & 10 & 10   & Yes \\
 NDW & 0.3 & 1  & 1 & 5 & 14    & Yes \\ 
 %NDW & 0.5 & 1  & 4 & 1 & 5 & 0.3 & 10  & Yes \\
 %NDW & 0.5 & 1  & 0.25 & 1 & 20 & 0.3 & 10  & No \\
 %NDW & 0.5 & 1  & 40 & 10 & 5 & 0.3 & 10  & Yes \\
 %NDW & 0.5 & 1  & 2.5 & 10 & 20 & 0.3 & 10  & No \\
 MW & 0.5 & 0.5  & 1 & 1 & 10  & Yes \\ 
 MW & 0.5 & 1  & 1 & 1 & 10   & Yes \\ 
 MW & 1 & 0.5  & 1 & 1 & 10   & No \\ 
 MW & 1 & 1  & 1 & 1 & 10   & Yes \\ 
 [1ex] 
 \hline 
\end{tabular}}
\caption{List of models and the 
initial conditions of our 2d spherical  simulations. Here $t_{\rm w}$ corresponds to the time the wind is active, which is related to the delay time between the merger and the collapse to a BH. $t_{\rm j}$ is the time the central engine is active. $L_{\rm j}$ corresponds to the isotropic jet luminosity. $\dot{M}_{\rm w}$  corresponds to the mass loss rate in the polar region of the wind. $\theta_0$ is the initial half-opening angle of the jet. All of our simulations have an inner radius of $1\times 10^{9}$cm and an outer radius of $6\times10^{10}$cm. We use an adaptive grid of size $l_{\rm r}=6\times 10^{10}$ cm and $l_\theta=\pi/2$ with $100\times 40$ initial cells and five levels of refinement resulting in a maximum resolution of $3.75\times 10^7$cm.  We also use common values for the  velocity of the wind $v_{\rm w}=0.3$ and  the jet Lorentz factor $\Gamma_{\rm j}=10$. For the density profile  of the wind we use a spherical wind (SW), a neutrino-driven wind (NDW) and a magnetized wind (MW). }
%\textcolor{magenta}{Maybe we can give names/acronyms to these models, and then refer to them in the Figure captions where snapshots are presented?}}
\label{table:data}
\end{table*}

\subsection{The wind medium in the pre-collapse, merged remnant}
\begin{figure}[t!]
\centering
\includegraphics[width=0.47\textwidth]{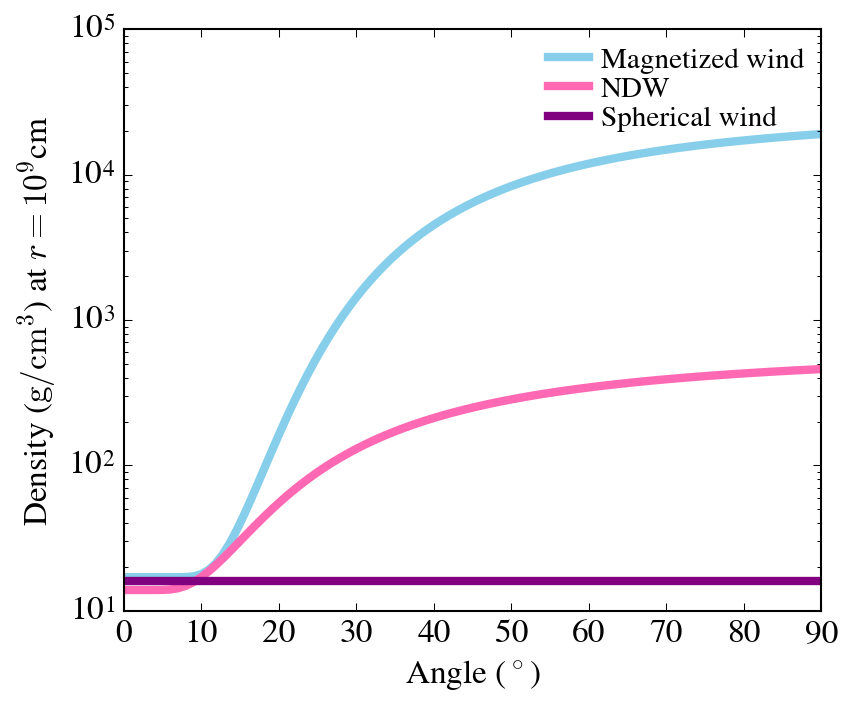}
\caption{ The angular structure of the density profiles  used in our simulations. The density is plotted at $r=10^9$cm, the inner boundary of the simulations. The three different wind prescriptions are  plotted and have a mass loss rate (at the polar region) of $\dot{M}_{\rm w} (\theta=0^\circ)=10^{-3}M_\odot$/s. The density profiles are derived from global simulations by \citet{2014MNRAS.443.3134P} for the NDW, and by \citet{2019ApJ...882..163J} for the magnetized wind. In this figure the angular radial profiles are derived by  fitting smooth functional forms to the numerical profiles.}
\label{fig:different_winds}
\end{figure}

Three different wind prescriptions   are simulated, which have different angular structures. The first is a constant uniform spherical wind \citep{ari14}. The second is a non-spherical latitudinal distribution in density and velocity \citep{ari17}, whose distribution is based on the global simulations of \citet{2014MNRAS.443.3134P}, accurately representing the neutrino-driven wind from the merger remnant. Additionally, we have included a density profile of a magnetized wind outflow  based on the simulations by \citet{2019ApJ...882..163J},  where the author performs general relativistic magneto-hydrodynamical simulations\footnote{We note that although the calculations of \citet{2019ApJ...882..163J} were performed for a 3 $M_\odot$ BH surrounded by 0.1 $M_\odot$ accretion disk, the resultant angular profiles are very similar to those by \citet{Siegel14} for a merger remnant. These wind profiles have the advantage that they have been calculated over many more dynamical times, due to the relative numerical simplicity, and thus better resemble a steady state wind solution.}. The  density profiles in both cases are similar, they are denser in the equator and progressively lighter in the polar region. The density as a function of angle is obtained by averaging the latitudinal profiles at the end of the simulations  \citep{2014MNRAS.443.3134P,2019ApJ...882..163J}.  The density of the wind is given by
\begin{equation}
\rho_{\rm w}(\theta)=\frac{\dot{M}(\theta)}{4\pi r^2 v_{\rm w}},
\end{equation}
 where $\dot{M}_{\rm w}(\theta)$ is the mass loss rate, $r$ is the radial coordinate, $\theta$ is the angular component, and $v_{\rm w}$ is the velocity of the wind, which is assumed to be constant.
For comparison, we plot the angular distributions of density in the three different winds in Figure~\ref{fig:different_winds}. The ratio of the density in the polar region ($\theta=0^\circ$) and the equator ($\theta=90^\circ$) is $\rho_{\rm eq}/\rho_{\rm pl}\approx10^2$ for the NDW and $\rho_{\rm eq}/\rho_{\rm pl}\approx 1.5\times 10^3$ for the magnetized outflow. We use two different global values for the mass loss rate, as prescribed in the polar region: $10^{-3}M_\odot$/s and $10^{-2}M_\odot$/s. This range is motivated by the range of values seen in magnetically-driven and neutrino driven wind calculations \citep{2014MNRAS.443.3134P,Siegel14,2017PhRvD..95h3005S,2018ApJ...860...64F,2019ApJ...870L..20N,2019ApJ...877L..40G}.

\begin{figure*}[t!]
\centering
\includegraphics[width=0.7\textwidth]{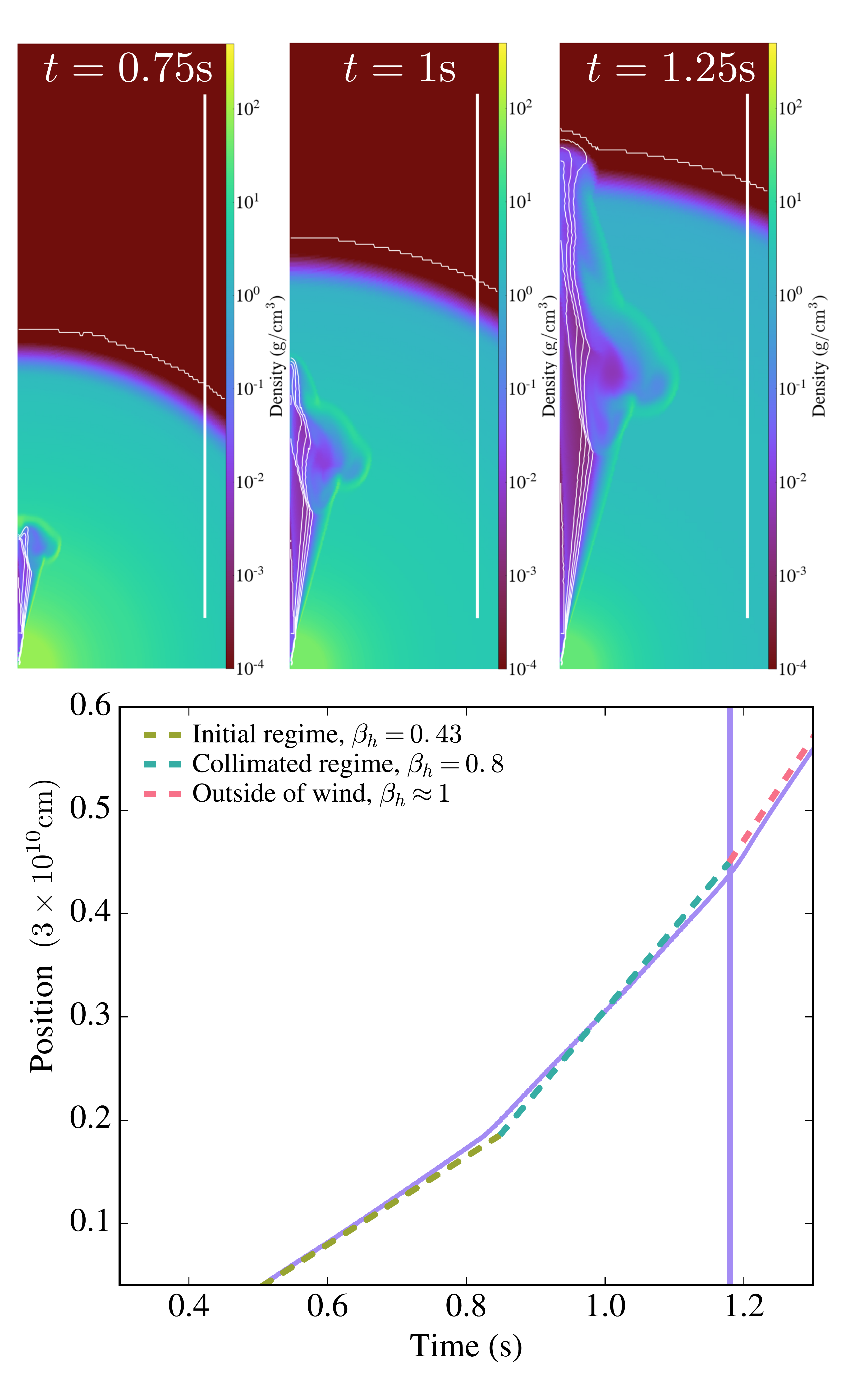}
\caption{ \emph{Top} panels: Density snapshots of the jet interacting with a spherical wind with $\dot{M}_{\rm w}=10^{-2}M_\odot$/s. The jet has a luminosity of $L_{\rm j}=10^{51}$erg/s, $\theta_0=10^\circ$ and $t_{\rm j}=1$s, while the wind is active for $t_{\rm w}=0.5$s before collapse. In this and all other simulations the wind starts at $t=0$ and is active for a time $t_{\rm w}$. After this, a jet is triggered, whose duration is $t_{\rm j}$.  The panels show the evolution of the jet at three different times and the white contours are lines of constant Lorentz factor. Shown in all cases is a $1.5\times 10^{10}$ cm scale bar. \emph{Bottom} panel: The temporal evolution of the position of the head of the jet (purple). Plotted in dashed lines are different  constant velocity regimes as estimated by the analytical formalism: initial expansion at constant $\tilde{L}$, the recollimation region with varying $\tilde{L}$ and the free expansion phase. The vertical line gives the time at which the jet breaks free from the wind region.
}
\label{fig:jet_evol}
\end{figure*}

\subsection{Evolution of the jet}
Initially, the jet is
unable to move the wind material at a speed comparable to its own and thus is  decelerated. As the jet propagates in the wind  a bow shock runs ahead of it, which both heats material
and causes it to expand sideways \citep{2002MNRAS.337.1349R, 2011ApJ...740..100B,2020A&A...636A.105S}. 
The parameter that controls the evolution of the jet interacting with the wind is \citep{2011ApJ...740..100B, ari17}:
\begin{equation}
\tilde{L}=\frac{\rho_{\rm j}h_{\rm j}\Gamma_{\rm j}^2}{\rho_{\rm w}\Gamma_{\rm w}^2},
\label{eq:ltilde}
\end{equation}
where $\rho$ is the density, $h$ is the enthalpy, $\Gamma$ is the Lorentz factor and the subscripts $j$, $w$ refer to the jet and the wind, respectively. 
 At a given time, the jet will have
evacuated a channel out to some location where it impinges on the pre-collapse wind, at a working surface advancing at velocity $\beta_{\rm h}$.  We balance the momentum fluxes at the working surface to obtain \citep{1989ApJ...345L..21B,2011ApJ...740..100B,ari17}
\begin{equation}
\label{eq:beta_h}
    \beta_{\rm h}=\frac{\beta_{\rm j}+\beta_{\rm w}\tilde{L}^{-1/2}}{1+\tilde{L}^{-1/2}},
\end{equation}
 where $\beta_{j}=v_{\rm j}/c$ is the initial velocity of the jet, as determined by $\Gamma$.
{ During propagation in the  wind, the head of the jet will initially expand at a constant velocity. This is because $\rho_{\rm j}/\rho_{\rm w}$ is independent of $r$, so that $\tilde{L}$ remains unchanged (Equation~\ref{eq:ltilde}). This can be seen in the expansion of the head of the jet during  the first $0.75$s (shown in the {\it bottom panel} of Figure~\ref{fig:jet_evol}), which  expands at a constant velocity. The surplus energy during this time is deposited within a cocoon surrounding the jet. As the pressure of the cocoon cavity increases, a recollimation shock is formed which minimizes the cross-sectional area of the jet (Figure~\ref{fig:jet_evol}). In this region, $\tilde{L}$ depends on the pressure build up within the cocoon region \citep[$P_{\rm c}$;][]{2011ApJ...740..100B}} 
\begin{equation}
\tilde{L}=\frac{4 P_{\rm c}}{\theta_0^2\rho_{\rm w}c^2}.   
\end{equation}

This transition  is clearly seen after $0.75$s in the evolution of the jet's head plotted  in  Figure~\ref{fig:jet_evol}. 

In order for the jet to be successful, the central engine needs to remain active when the jet breaks out of the wind. By considering the time at which the head of the jet reaches the outer edge of the wind, the condition for a successful SGRB jet can be written as:
\begin{equation}
\label{eq:successful_jet}
    \beta_{\rm h}> \beta_{\rm h,c} = \beta_{\rm w}\left(1+ {t_{\rm w}\over t_{\rm j}}\right) \;,
\end{equation}
where $\beta_{\rm w}=v_{\rm w}/c$. This condition is clearly satisfied in the simulation shown in  Figure~\ref{fig:jet_evol}. At the time the jet head breaks out from the wind ($t\approx 1.2$s), we required  $\beta_{\rm h}> \beta_{\rm h,c}\approx 0.45$, which is clearly satisfied and the jet remains active.  In Section \ref{sec:swcond}  we investigate the applicability of condition \ref{eq:successful_jet} for both spherical and non-spherical winds.

\subsection{Winds, jet dynamics and successful jets}\label{sec:swcond}
\begin{figure*}
\centering
\includegraphics[width=0.78\textwidth]{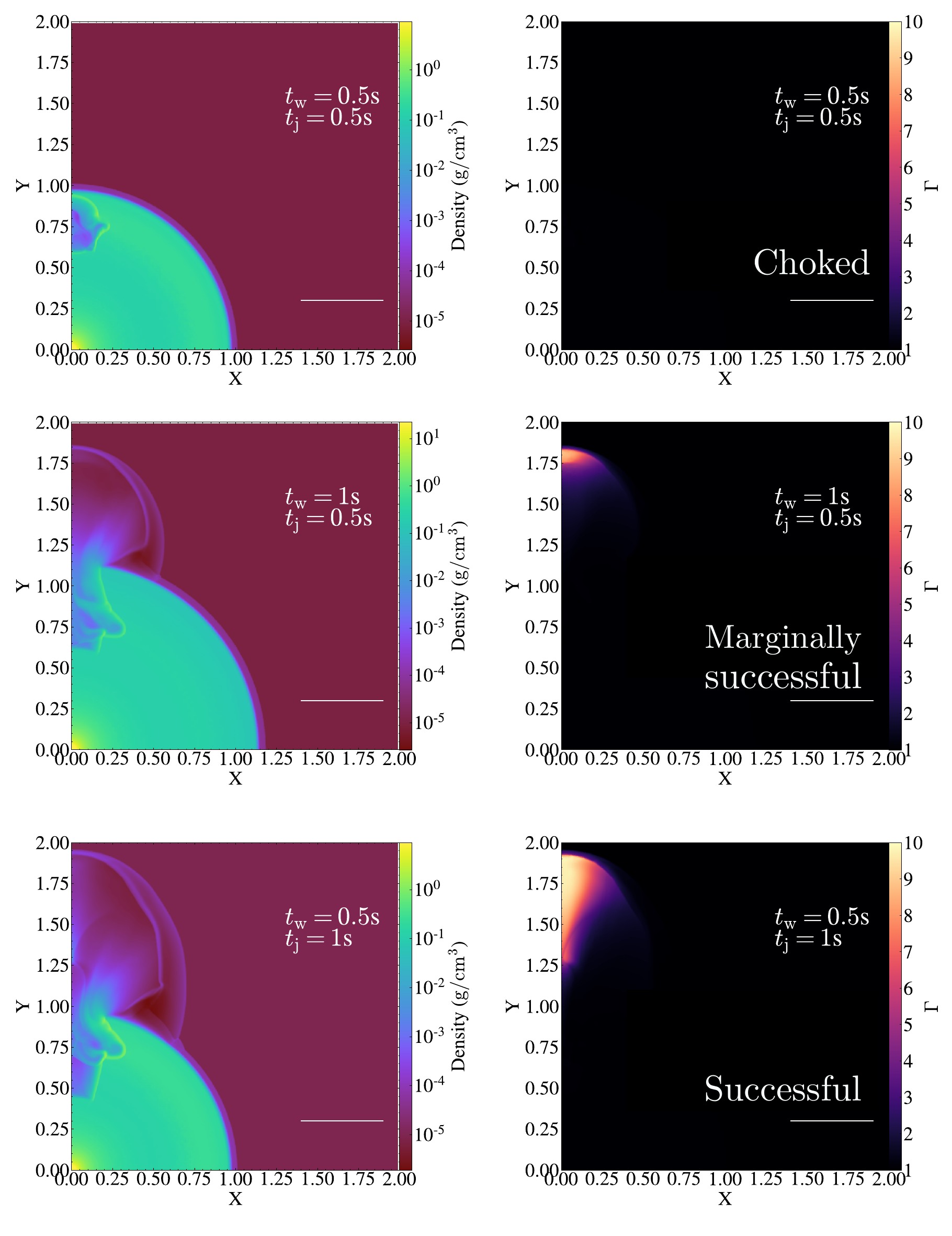}
\centering
\caption{Simulations showing how the  expansion of the jet is affected by the properties of the wind through which it propagates and in particular, its pre-collapse duration, $t_{\rm w}$. Three illustrative cases are depicted of simulations of jets propagating within a spherical wind with  a mass loss rate of $\dot{M}_{\rm w}=10^{-2} M_\odot$/s and $v_{\rm w}=0.3c$.  The {\it left} panels show the density while {\it right} panels show the Lorentz factor. The {\it top} and {\it bottom} panels  correspond to the evolution of the jet at a time of 2.75s, while the {\it middle}  panel is a snapshot  at a time of 3.25s. Shown in all cases is a $1.5\times 10^{10}$ cm scale bar.   \textit {Top} Panel:  A choked jet with $L_{\rm j}=10^{50}$erg/s, $t_{\rm w}=0.5$, $t_{\rm j}=0.5$. \textit {Middle} Panel:  A marginally successful jet (i.e., $\beta_{\rm h}\approx \beta_{\rm h,c}$)  with $L_{\rm j}=10^{51}$erg/s, $t_{\rm w}=1$ and $t_{\rm j}=0.5$. \textit {Bottom} Panel: A successful jet with $L_{\rm j}=10^{51}$erg/s, $t_{\rm w}=0.5$ and $t_{\rm j}=1$.} 
\label{fig:sims_all_cases_cw}
\end{figure*}

The shocks responsible for producing a sGRB must arise after the relativistic jet has broken free from  the pre-collapse wind. The majority of compact mergers, with the exception of those involving a black hole, will not collapse immediately and a dense wind will thus remain to impede the advance of the jet. As stated by the condition \ref{eq:successful_jet}, a sGRB is likely to be produced if the jet triggered by the post-collapse accretion maintains its power for longer than it takes the jet to reach the edge of the wind. If this is not the case, 
as shown in the {\it top} panel in Figure~\ref{fig:sims_all_cases_cw}, the jet will be choked. On the other hand, if the jet continues to be active and  emerges from the wind, the sudden and drastic density drop at the outer edges allows the jet head to accelerate to velocities close to the speed of light\footnote{ In the case depicted in the {\it} bottom panel of Figure~\ref{fig:jet_evol}, the final Lorentz factor is approximately $9$, which is 90\% of the initial Lorentz factor.},  as illustrated in the {\it middle} and {\it bottom} panels in Figure~\ref{fig:sims_all_cases_cw}. In the   {\it middle} panel $\beta_{\rm h}\approx \beta_{\rm h,c}$, while in the {\it bottom} panel  $\beta_{\rm h}\gg \beta_{\rm h,c}$.

\begin{figure}[t!]
\centering
\includegraphics[width=0.45\textwidth]{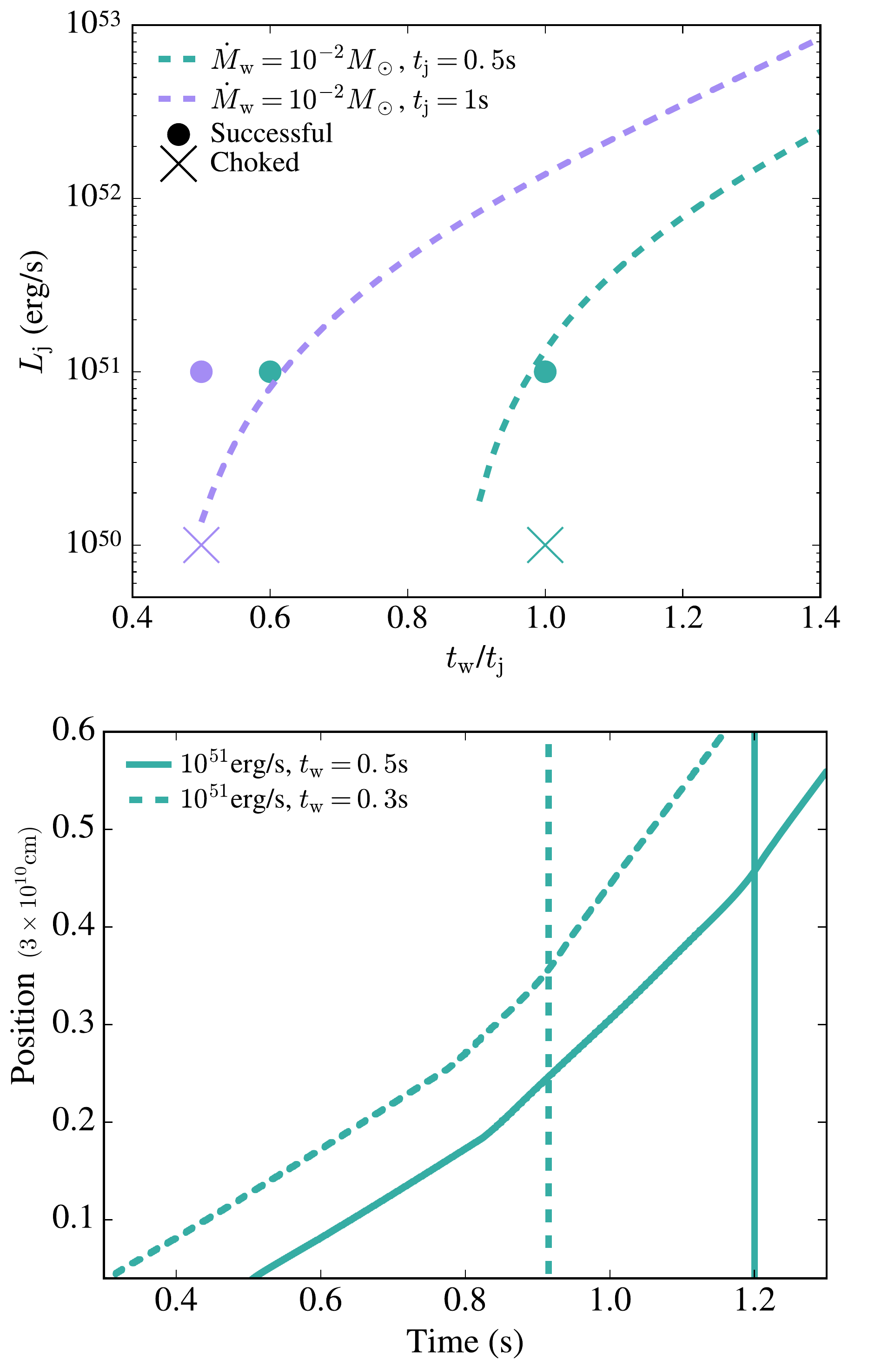}
\caption{ \emph{Top} panel: The critical  power needed in order to produce  a successful sGRB for  jets with varying $t_{\rm w}/t_{\rm j}$. This condition is derived from equation~\ref{eq:successful_jet} assuming $\tilde{L}$ is constant. All cases are for a spherical, isotropic wind with a mass loss rate of $\dot{M}_{\rm w}=10^{-2}M_\odot$/s.  The unsuccessful jet simulations are marked with crosses and the dots represent successful cases. \emph{Bottom} panel: The position of the head of the jet as a function of time for two of the teal
cases plotted in the \emph{top} panel. The corresponding vertical lines show when the jet breaks out of the wind. The dashed  lines correspond to a simulation with $t_{\rm w}=0.3$s, while the solid line is for a simulation with $t_{\rm w}=0.5$s. The transition to the recollimation regime happens at $0.83$s ($0.78$s) for the $t_{\rm w}=0.5$s  ($t_{\rm w}=0.3$s) case.
}
\label{fig:lum_crit}
\end{figure}

The condition $\beta_{\rm h}>\beta_{\rm h,c}$ can be used to derive the critical power needed for a jet to be successful, as argued by \citet{ari17}. This critical luminosity estimate is shown in Figure~\ref{fig:lum_crit} in the context of our jet simulations with varying $t_{\rm w}/t_{\rm j}$ and expanding in a medium with  $\dot{M}_{\rm w}=10^{-2}M_\odot$/s. The dots in Figure~\ref{fig:lum_crit} correspond to successful jets and the crosses are choked jets. 
As expected, the ratio $t_{\rm j}/t_{\rm w}$ is critical in determining successful  breakout of the expanding jet.

\begin{figure}
\centering
\includegraphics[width=0.33\textwidth]{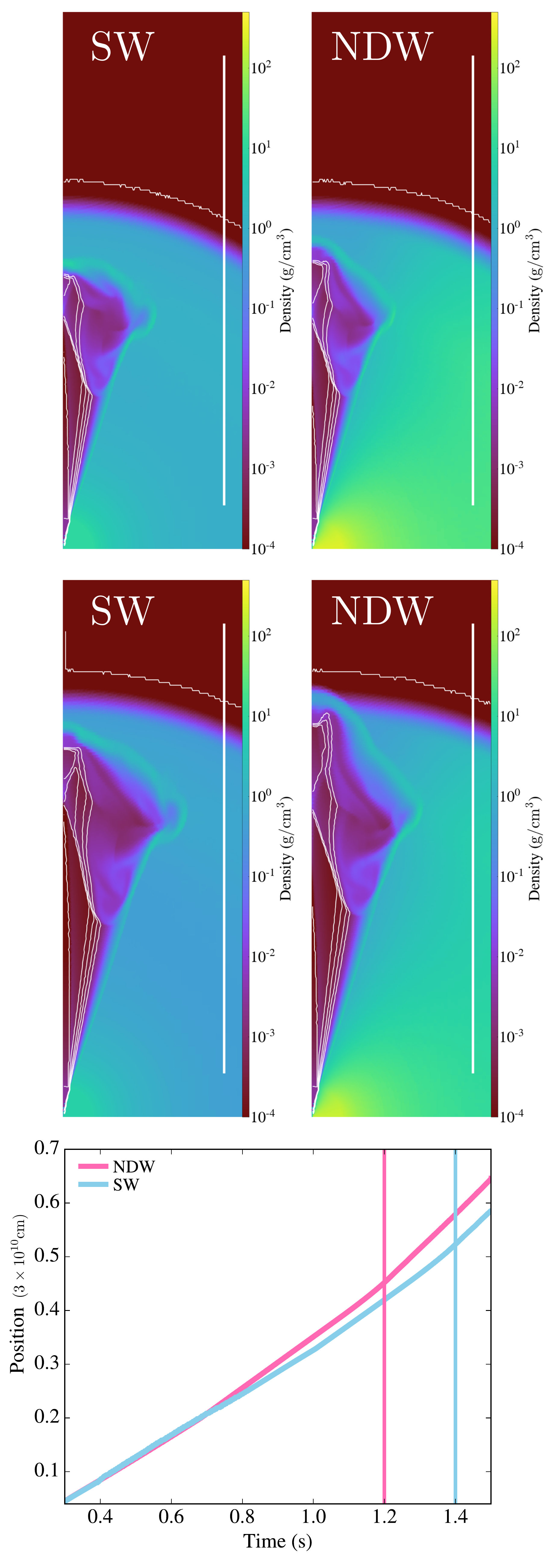}
\caption{ The evolution of two identical jets propagating in  wind environments with different structure. The simulations are for $L_{\rm j}=10^{50}$erg/s, the mass loss rate in the polar region is $\dot{M}_{\rm w}=5\times10^{-3}M_\odot$/s, $t_{\rm w}=0.3$s, and $t_{\rm j}=1$s, and $\theta_0=14^\circ$. Here SW refers to a spherical wind while NDW refers to the neutrino-driven wind (NDW). Shown is a $1.5\times 10^{10}$ cm scale bar.  \emph{Top} panels: Comparison between the evolution of the jets at 
$t=1$s. The contour lines represent lines with equal Lorentz factors. Shown in all cases is a $1.5\times 10^{10}$ cm scale bar.  \emph{Middle} panels: Same as the \emph{top} panels but for $t=1.25$s. \emph{Bottom} panel: The position of the head of the jet for the two cases depicted above. The vertical lines gibe the time at which  the jet breaks from the wind. The transition to the collimated regime occurs at $0.74$s for the NDW and at $1$s for the SW. }
\label{fig:jet_zoom}
\end{figure}

{ The critical  condition for the minimum jet power in Figure~\ref{fig:lum_crit}  has been derived assuming that $\tilde{L}$ is constant. This is generally a valid assumption when the jet expands in a  $1/r^2$ wind medium and experiences little recollimation. The applicability of this approximation can be seen by contrasting the evolution of  the  two jets depicted in  the {\it bottom panel}  in  Figure~\ref{fig:lum_crit}. One of them (solid line) experiences significant recollimation before reaching the edge of the jet, while the other one (dashed line) escapes before the cocoon pressure is able to significantly alter the jet's initial structure. We note that although both jets are successful, the condition plotted in the {\it top} panel of Figure~\ref{fig:lum_crit} places the simulation with $t_{\rm w}/t_{\rm j}=1$ (solid line in the {\it bottom} panel) below the line. In this case, recollimation permits the jet to expand at a faster rate than the one predicted using the constant $\tilde{L}$ assumption. We thus caution the reader that such a constraint should be used as a conservative limit for the necessary jet power. 

Additionally, the general formalism used to derive condition given by Eq.~\ref{eq:successful_jet} makes use of spherical symmetry and as such its applicability is less rigorous when the jet expansion takes place within a non-spherical mass distribution. Figure~\ref{fig:jet_zoom} compares the evolution of two identical jets propagating in two different wind environments. The density profiles in both cases are similar in the polar region but  NDW's profile is progressively denser as it approaches the equatorial region (Figure~\ref{fig:different_winds}).
When the jet propagates in the NDW, the cocoon region is significantly more confined due to the increase in the surrounding pressure. As a result, recollimation occurs earlier in the jet's propagation when compared to the spherical case and causes the jet to propagate at a faster speed. It is thus important to note that in the realistic wind cases considered in this paper, condition given by Eq.~\ref{eq:successful_jet} provides a rather robust limit for the minimum luminosity require to produce a successful jet.
}

As demonstrated here, two dimensional simulations have uncovered some dynamical properties of relativistic flows unanticipated by analytical models, but we caution that there are still  some key questions that they cannot tackle. A feature of the jet-wind interaction  is that there will surely be some mass entertainment and maybe more importantly, there can be an exchange of linear momentum with the surrounding material. Because of this, we cautioned that higher resolution is needed because even a tiny mass fraction of baryons loading down the jet could severely limits the attainable Lorentz factor (here assumed to be only 10 for numerical convenience). Additionally, the symmetry-breaking involved
in transitioning from two to three dimensions is crucial for understanding the nonlinear development of instabilities, leading to qualitatively new phenomena \citep{2008A&A...488..795R,2010A&A...520L...3M,2013ApJ...767...19L,2018MNRAS.479..588G,2018MNRAS.473..576G,2019MNRAS.490.4271M,2020arXiv200602466G}. Another topic which seems ripe for a more sophisticated treatment concerns the possibility  that the initial jet is a magnetically confined configuration, whose collimation properties are not as drastically modified by the distribution of the external material \citep{2012ApJ...757...16M,2016MNRAS.456.1739B,2016MNRAS.461L..46T,2019ApJ...877L..40G,2020MNRAS.495.3780N,2020arXiv200711590G,2020arXiv200909714N,2020arXiv200910475M}.

\begin{figure}
\centering
\includegraphics[width=0.47\textwidth]{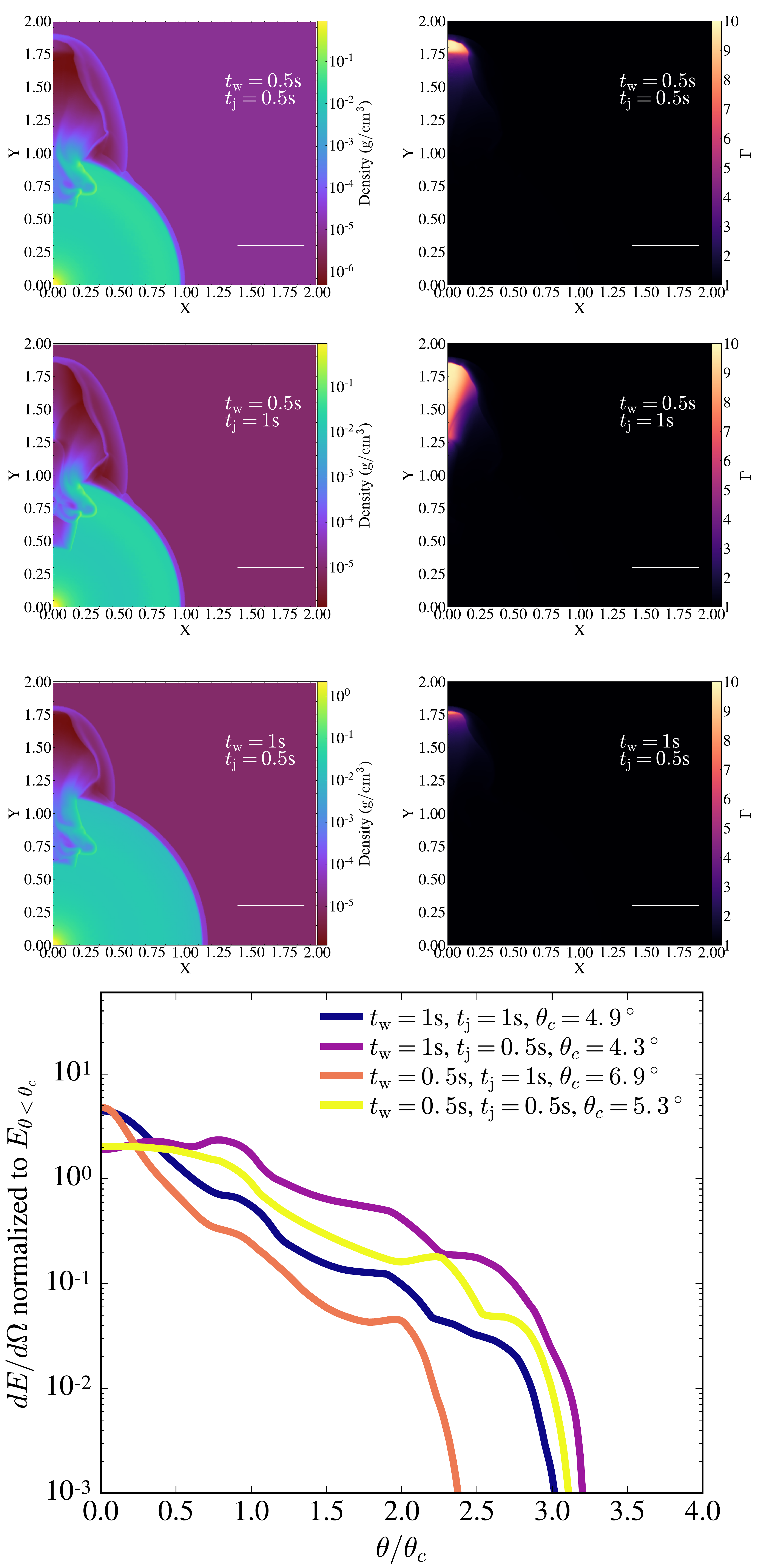}
\caption{ \emph{Top} panels: Density  ({\it left}) and Lorentz factor ({\it right}) profiles of simulations of the interaction of a relativistic jet with a spherical wind. The bar corresponds to $1.5\times 10^{10}$cm. The luminosity is $L_{\rm j}=1\times10^{50} \rm{erg/s}$, the initial Lorentz factor $\Gamma=10$, and the initial half-opening angle $\theta_0=10^\circ$. The wind has an $\dot{M_{\rm w}}=10^{-3} M_\odot$/s in the polar region and $v_{\rm w}=0.3c$. Different simulations assume different collapse times and jet lifetimes. The simulations were run up to $4$s. The \emph{top} and \emph{middle} snapshots were taken after $2.75$s while the \emph{bottom} one was taken after $3.25$s.  \emph{Bottom} panel: Energy per unit angle of the resulting jet. The time is the same as the above panel.  For $t_{\rm w}=1$ and $t_{\rm j}=1$, the time is $3.25$s. The energy is normalized to the total energy in the core of the jet. } 
%\end{minipage}
\label{fig:sims_cw}
\end{figure}

\begin{figure}
\centering
\includegraphics[width=0.48\textwidth]{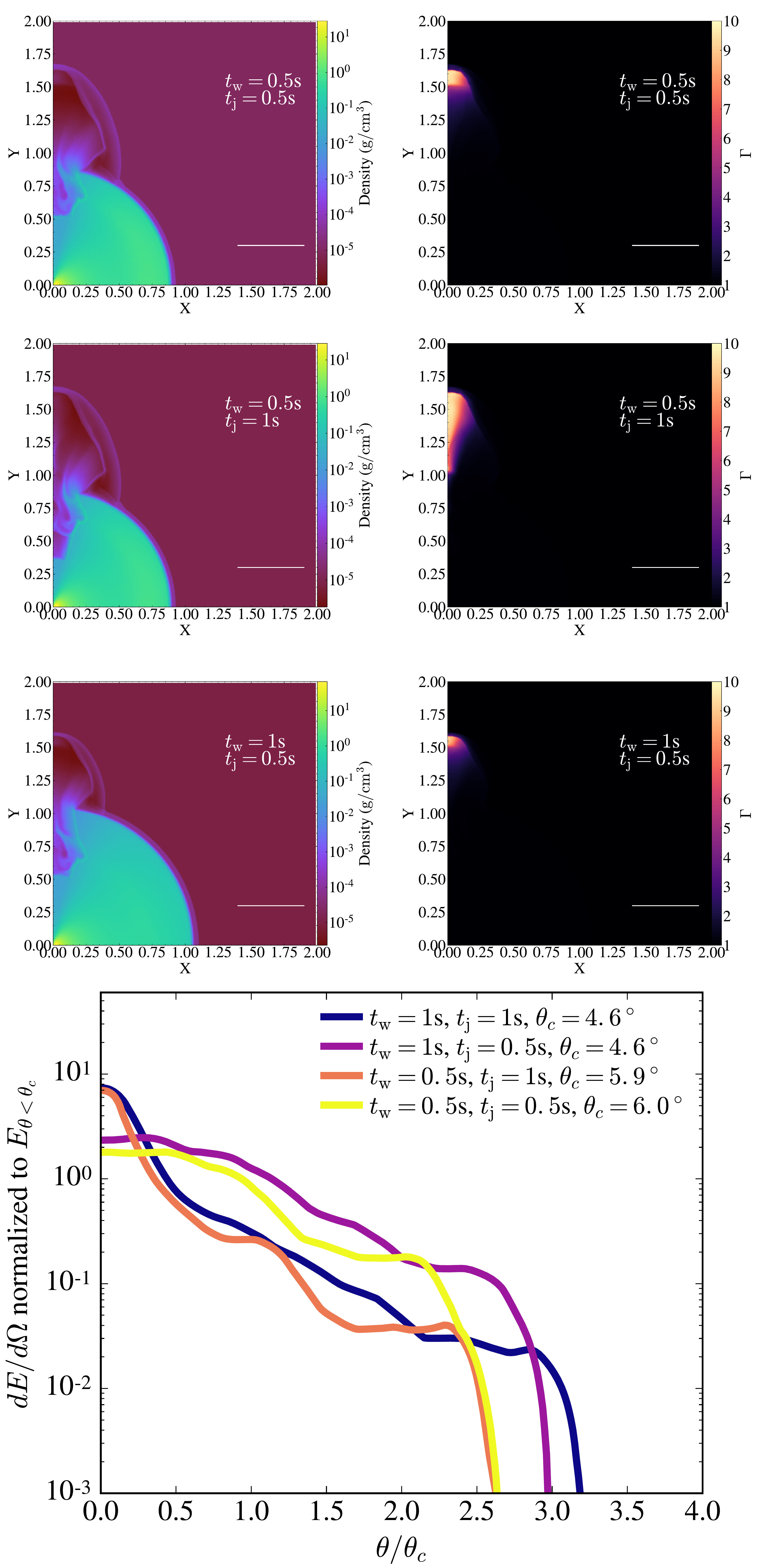}
\centering
\caption{ \textit{Top} panels: Density  ({\it left}) and Lorentz factor ({\it right}) profiles of simulations of the interaction of the relativistic jet with a neutrino-driven wind. Shown is a $1.5\times 10^{10}$ cm scale bar. The properties of the jet are the same as in Fig~\ref{fig:sims_cw}. The {\it top} and {\it middle} panels show snapshots after  $2.5$s while $3$s for the {\it bottom} panel. The wind has a mass loss rate of $\dot{M}_{\rm w}=10^{-3} M_\odot$/s in the polar region and a velocity of $v_{\rm w}=0.3c$. 
\textit {Bottom} panel: Energy per unit angle of the jet after its propagation. The time is the same as the above panel, for $t_{\rm w}=1$ and $t_{\rm j}=1$, the time is $3$s, and for $t_{\rm w}=0.3$ and $t_{\rm j}=1$, the time is $2.25$s. The energy is normalized to the energy in the core.} 
\label{fig:sims_ndw}
\end{figure}

\begin{figure}
\centering
\includegraphics[width=0.47\textwidth]{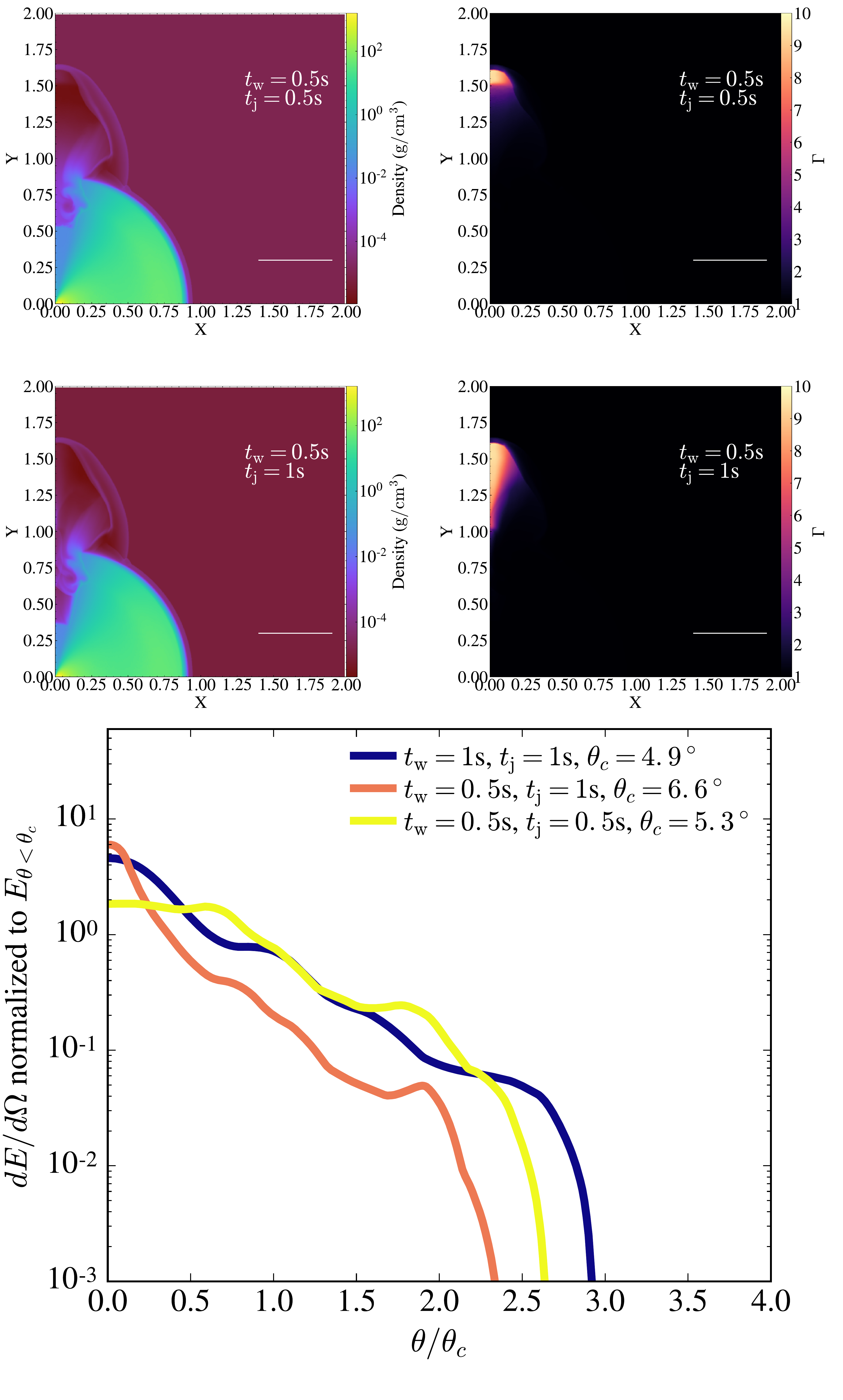}
\centering
\caption{ \textit {Top} panels: Density  ({\it left}) and Lorentz factor ({\it right}) profiles of simulations of the interaction of a relativistic jet with a magnetized wind based on \citet{2019ApJ...882..163J}. Shown is a $1.5\times 10^{10}$ cm scale bar. The properties of the jet are the same as in Fig~\ref{fig:sims_cw}. The wind has a mass loss rate of $\dot{M}_{\rm w}=10^{-3} M_\odot$/s in the polar region and $v_{\rm w}=0.3c$.  The {\it top} and {\it middle} Panels correspond to  a simulation time of $2.5$s. The simulation corresponding to $t_{\rm w}=1$s, $t_{\rm j}=0.5$s is not shown as the wind in that case is dense enough to choke the jet, rendering the sGRB unsuccessful. 
\textit {Bottom} panel: Energy per unit angle of the jet resulting from the simulations. The time is the same as the above panel. For $t_{\rm w}=1$ and $t_{\rm j}=1$, the time is $3.25$s. The energy is normalized to the energy in the core.} 
\label{fig:sims_aw}
\end{figure}

\subsection{Jet structure}
{ We perform different simulations, for a given wind profile, by altering  the lifetime of the wind ($t_{\rm w}$), which is directly related to the time it takes the merger remnant to collapse to a black hole. We also changed the duration of the jet ($t_{\rm j}$), which is the characteristic time the central engine is active and is broadly related to the duration of the event. We show our results in  Figures \ref{fig:sims_cw}, \ref{fig:sims_ndw} and \ref{fig:sims_aw}, where we plot the density and Lorentz factor distributions of jets propagating through a spherical wind, a neutrino-driven wind and  a magnetically ejected wind, respectively. The energy per unit solid angle at the end of the simulation is estimated as in \citet{2015ApJ...813...64D}.

The interaction of the jet with the wind  will result in an angular redistribution of the jet's energy, which  can naturally give rise to a different afterglow light curve than the one from the original top-hat structure \citep{2017MNRAS.471.1652L,2018ApJ...866....3D,2018MNRAS.475.2971B,2018MNRAS.481.1597G,2018MNRAS.478..733L,2018MNRAS.473L.121K,2018arXiv180806617V,2019ApJ...870L..15L,2019A&A...628A..18S,2020ApJ...898...59L,2020arXiv200602466G}. In what follows, we explore  how the properties of the pre-collapse wind affect the structure of a jet propagating through it. 

Common to all calculations is the altering of $t_{\rm w}$ and $t_{\rm j}$.  Also shown is the final angular distribution  of the energy in the relativistic jet. In all simulations we define the core angle of the jet, after its propagation, as the angle where the Lorentz factor of the jet decreases by a factor of 2 from its initial value. The resulting core angles range from $4.5^\circ$ to $7^\circ$, which are similar to the values quoted in the literature \citep{2018NatAs...2..751L,2018A&A...613L...1D,2018MNRAS.478L..18T, 2019A&A...628A..18S,2019Sci...363..968G, 2020ApJ...898...59L,2020MNRAS.495.3780N}. The final angular distribution of Lorentz factor is found to be well represented by a  Gaussian distribution. In all cases, the jet, which is originally a top-hat, spreads laterally thus resulting in a  structured jet. }

In addition to the density of the wind, which depends primarily on the mass loss rate, the ratio $t_{\rm w}/t_{\rm j}$ has a decisive effect on the appearance of a jet propagating through it. This is because it determines the time the jet resides within the interaction region (as governed by $\tilde{L}$) which in turn regulates the amount of  relativistic  material that is shocked. The importance of this ratio can be clearly seen in the {\it Bottom} panels of Figures~\ref{fig:sims_cw}, \ref{fig:sims_ndw} and Figure~\ref{fig:sims_aw}. If the delay time is larger or comparable to the duration of the jet, the time it takes for the jet to break free from the wind is augmented. In this case, the afterglow emission would be dominated by the emission of the laterally spreading  relativistic  material, which is located  at larger angles relative to the rotation axis of the merger remnant. On the other hand, if the jet produced by the accretion onto the black hole maintains its energy for much longer than it takes the jet head to reach the  edge of the wind, the core of the relativistic jet would contain substantially more energy than the off-axis  material, so that it is likely to dominate the afterglow flux even after expanding for a longer time. The detection of varying afterglow signatures would be a test of the neutron star merger model; and the precise inference of the angular structure of the jet may help constrain the  properties of the wind  and the lifetime of the merger remnant.

\begin{figure}
\centering
\includegraphics[width=0.47\textwidth]{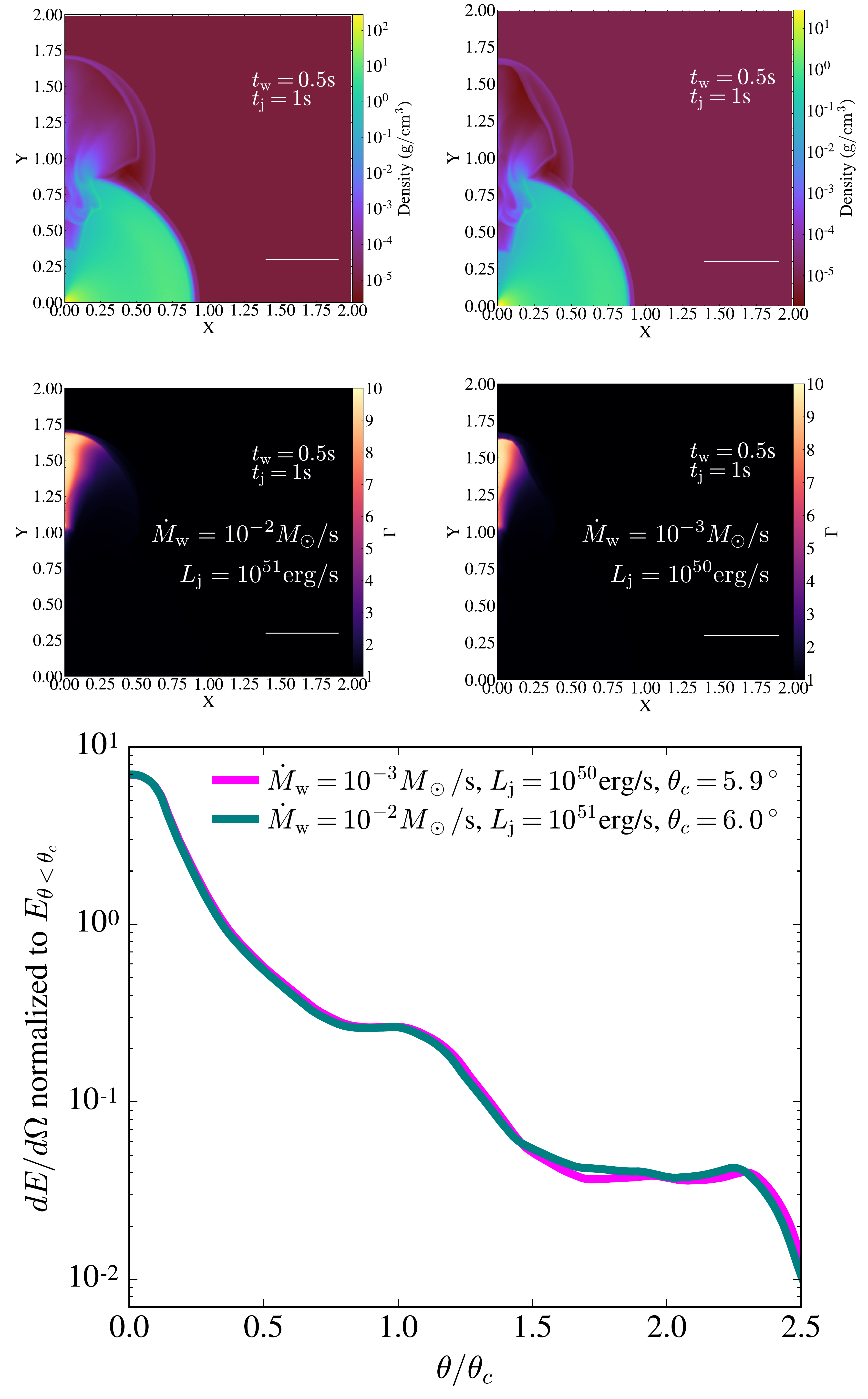}
\centering
\caption{\textit{Top} and \textit{middle}  panels: Density snapshots with Lorentz factor for simulations that have the same $\tilde{L}$. On the {\it left}, the wind has a mass loss rate in the equator of $\dot{M}_{\rm w}=10^{-2} M_\odot$/s and the jet has a luminosity of $L_{\rm j}=10^{51}$erg/s. On  the {\it right}, the wind has a mass loss rate in the equator of $\dot{M}_{\rm w}=10^{-3} M_\odot$/s and the jet has a  luminosity of $L_{\rm j}=10^{50}$erg/s. The  density profile used here is the  neutrino-driven wind. Shown is a $1.5\times 10^{10}$ cm scale bar. The jets are shown at $t=2.5$s. {\it Bottom} Panel: Energy per unit angle of the jet after its propagation.  The energy is normalized to the energy in the core. } 
\label{fig:sims_ndw_mdot}
\end{figure}

The strength of the baryon loaded wind is also a key parameter.  If the mass loss rate is increased to $\dot{M}_{\rm w}=10^{-2}M_\odot$/s in the polar region and the luminosity is kept at $L_{\rm j}=10^{50}$erg/s, the wind would choke the jet, rendering the sGRB unsuccessful. The jet, in this environment  can become successful if its power is increased, as governed by $\tilde{L}$ \citep{2011ApJ...740..100B}.   In Figure~\ref{fig:sims_ndw_mdot}, we show the effects of altering the mass loss rate and the luminosity of the wind but leaving  $\tilde{L}$. unchanged. As expected, the evolution  of the jet remains unchanged and the simulation outputs in these two cases look almost identical.

\begin{figure}
\centering
\includegraphics[width=0.43\textwidth]{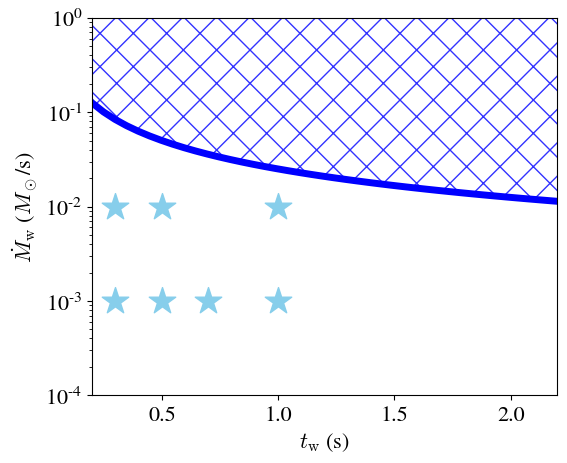}
\caption{ Distribution of mass loss rate and wind duration time. The blue stars represent the values for the simulations performed in this study. The blue line represents the combination of mass loss rate and $t_{w}$ that give an ejecta mass of $M_{\rm ejecta}=0.025 M_\odot$ \citep{2017Natur.551...80K}. Above this line, the parameter space is not permitted by observations of the blue kilonova.} 
%\end{minipage}
\label{fig:mejecta}
\end{figure}

\begin{figure}[t!]
\centering
\includegraphics[width=0.5\textwidth]{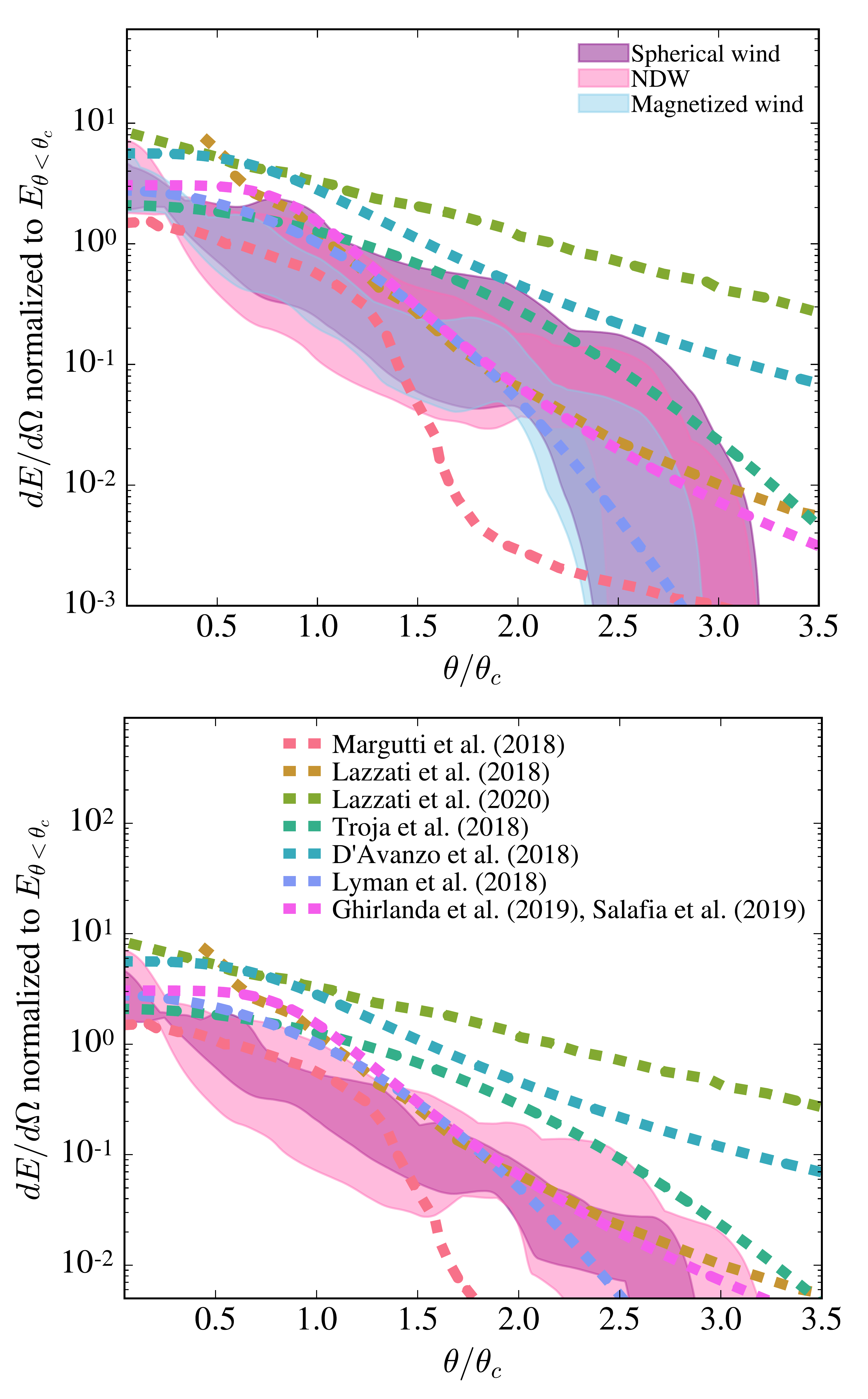}
\caption{Comparison of different jet profiles used to model the emission of GW170817 with those obtained from our simulations. \emph{Top} panel: Jet profiles for the various winds at a fixed jet luminosity of $10^{50}$erg/s but varying $t_{\rm w}/t_{\rm j}=0.3,\ 0.5,\ 1,\ 2$.  The dotted lines show the jet profiles inferred for GW170817 and   taken from \citet{2018PhRvL.120x1103L, 2018ApJ...856L..18M,2018NatAs...2..751L,2018A&A...613L...1D,2018MNRAS.478L..18T,2019Sci...363..968G,2019A&A...628A..18S,2020ApJ...898...59L}. \emph{Bottom} panel: Jet profiles resulting from the interaction of a jet with a NDW and a spherical wind with $\dot{M}_{\rm w}=10^{-2}M_\odot$/s in the polar region. The jet's luminosity is $10^{51}$erg/s, and $t_{\rm w}/t_{\rm j}=0.3,\ 0.5,\ 1,\ 2$ for the NDW, while $t_{\rm w}/t_{\rm j}= 0.5,\ 1,\ 2$ for the SW.} 
\label{fig:profiles}
\end{figure}

 \begin{figure}[ht!]
\centering
\includegraphics[width=0.45\textwidth]{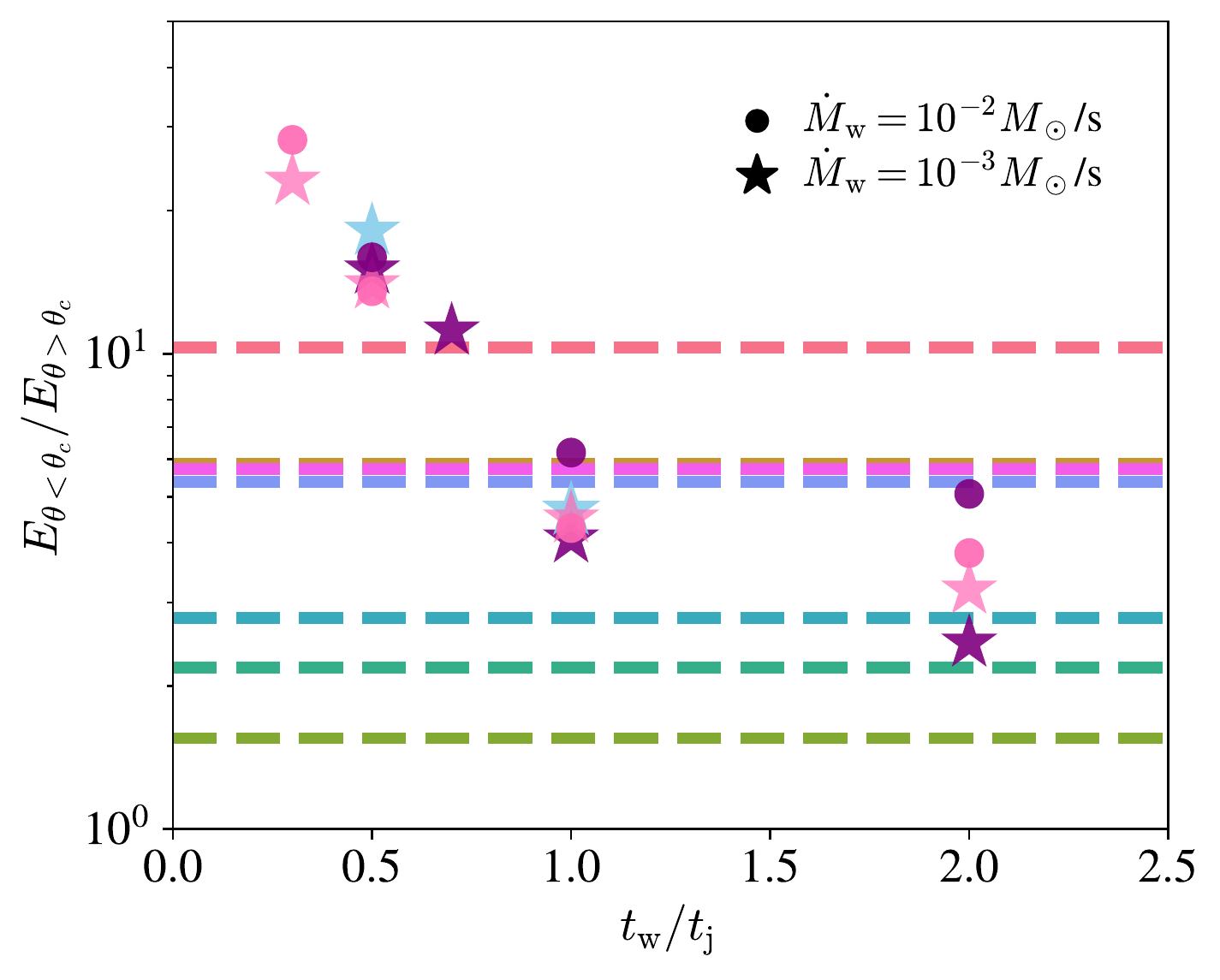}
\caption{Comparison of the relativistic energy content  in the jet's core, $E_{\theta < \theta_{\rm c}}$, with that stored in the wings, $E_{\theta > \theta_{\rm c}}$.  This ratio is plotted as a function of $t_{\rm w}/t_{\rm j}$ for the three different wind profiles: NDW (pink); spherical wind (purple) and magnetized wind (blue).  The stars represent simulations with $\dot{M}_{\rm w}=10^{-3}M_{\odot}$/s in the polar region and $L_{\rm j}=10^{50}$erg/s while the circles represent  simulations with $\dot{M}_{\rm w}=10^{-2}M_{\odot}$/s in the polar region and $L_{\rm j}=10^{51}$erg/s. The constant lines represent the same energy ratio taken from GW170817 jet simulations \citep{2018PhRvL.120x1103L, 2018ApJ...856L..18M,2018NatAs...2..751L,2018A&A...613L...1D,2018MNRAS.478L..18T, 2019Sci...363..968G,2019A&A...628A..18S,2020ApJ...898...59L},  where the color scheme for the various models is the same as in Figure~\ref{fig:profiles}.}
\label{fig:ratio}
\end{figure}

\section{Relevance to GW170817}
Several groups \citep{2018PhRvL.120x1103L, 2018ApJ...856L..18M,2018NatAs...2..751L,2018A&A...613L...1D,2018MNRAS.478L..18T,2019Sci...363..968G, 2019A&A...628A..18S,2020ApJ...898...59L} have studied the origin of the  afterglow emission of GW170817 and concluded that it can be explained by invoking a model where the sGRB was successful and the observer lies off axis to the jet. A common feature  of all models is the need for significant amount of energy at larger angles, which as we have argued here can  be a natural consequence of the  interaction of the jet with the pre-collapse wind. In this section we endeavor to compare our simulation results  to the jet models constructed for GW170817. 

In Figure~\ref{fig:mejecta} we plot the range of wind parameters  used in our study and compare them with the constraints derived by the presence of the blue kilonova. The  blue line represents the limit marked by the total mass derived  to produce the blue component of the kilonova  \citep{2017Natur.551...80K}. We note that the total mass ejected by the wind in our simulations  is below this value.

Various groups inferred different energy distributions for the jet. In Figure~\ref{fig:profiles} we compare their best fit models for the jet  profiles with those obtained from our simulations. The results from our simulations are in broad agreement with the energy  distributions derived from afterglow observations.
We thus conclude that the structure of an initially top-hat jet can be modified by its interaction with the  pre-collapse wind  and, after the jet emerges from this region, can have a structure that closely resembles the one deduced for GW170817. So in these models the $\gamma$-rays would be restricted to a narrow beam, even though outflow with a more moderate Lorentz factor, which is relevant to the afterglow emission, is spread over a wider range of angles. 

While the properties of the pre-collapse wind have an important effect on the appearance of a jet propagating through it (Figure~\ref{fig:profiles}), our calculations suggest that $t_{\rm w}/t_{\rm j}$ is the essential parameter that   controls how much relativistic energy is distributed at large angles.

Figure~\ref{fig:ratio} illustrates the effect of varying $t_{\rm w}/t_{\rm j}$ for jets propagating within the three different wind profiles we have considered in this study and the different mass loss rates. It shows how the ratio of the energy contained in the core of the jet to that residing outside it increases as $t_{\rm w}/t_{\rm j}$ augments.  As  argued above, we also see that changes in the mass loss rate  have a less dramatic  effect when compared to variations in $t_{\rm w}/t_{\rm j}$.

In the case of a successful break-through,  the resultant jet structure could result in an afterglow signature similar to that observed in GW170817 if the time it took for the merger remnant to collapse is similar to the observed duration of the event. An upper limit for $t_{\rm w}$ can be derived by requiring a successful jet  \citep{ari17}.   A successful jet can   be produced if the central engine remains active for a time longer than the time it takes for the jet to break through the wind:
\begin{equation}
t_{\rm w} \lesssim  t_{\rm j} \frac{\beta_{\rm h}-\beta_{\rm w}}{\beta_{\rm w}},
\label{eq:tw}
\end{equation}
where $\beta_{\rm w}=v_{\rm w}/c$, $\beta_{\rm h}=v_{\rm h}/c$ and  the subscript $h$ referring to the head of the jet.  This condition is derived using the evolution of the working surface,  where the velocity of the head of the jet is given by Equation~\ref{eq:beta_h}.

A lower limit for $t_{\rm w}$ can be found using Figure~\ref{fig:ratio}, which shows that in order for our models to explain the afterglow of GW170817, a long interaction with the wind is required.  In particular, in order to be in agreement with the most recent afterglow models, we require $t_{\rm w}/t_{\rm j}\gtrsim 0.7$. At smaller $t_{\rm w}/t_{\rm j}$, the core of the jet carries the bulk  of the energy and, as such, it would closely resemble a top-hat model which is inconsistent with current observations. Models that have more extended wings carrying a larger amount of energy thus require longer interaction times between the jet and the wind. Using $t_{\rm j}=2 \pm 0.5$s \citep{2017ApJ...848L..14G},  we derive the following stringent limit $t_{\rm w}\gtrsim 1.05 {\rm s}$. 

\begin{figure}
\centering
\includegraphics[width=0.43\textwidth]{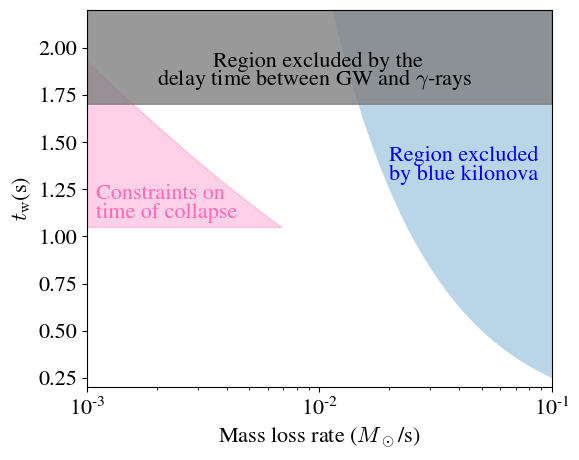}
\caption{ Constraints on $t_{\rm w}$  as a function of $\dot{M}_{\rm w}$. The upper limit is derived using Eq.~\ref{eq:tw}. We use  $t_{\rm j}=2 \pm 0.5$s \citep{2017ApJ...848L..14G} and the following  ranges obtained by \citet{2020ApJ...898...59L} for GW170817: $5\times 10^{48}-10^{50}$erg (jet energy) and $9-20^\circ$ (initial jet opening angle), which we use to obtain the isotropic luminosity. For consistency  we use $10^2<\Gamma<10^3$, yet its exact value does not meaningfully alter the value of $\beta_{\rm h}$ provided that the flow is highly relativistic. The grey region is excluded by the delay time between the gravitational wave (GW) and the $\gamma$-ray signal \citep{2017PhRvL.119p1101A,2017ApJ...848L..14G}. The blue region is the one excluded by  observations, which is also plotted in Figure~\ref{fig:mejecta}.} 
%\end{minipage}
\label{fig:tw_constrain}
\end{figure}

The constraint given above can be combined  with the  successful sGRB requirement given by Equation~\ref{eq:tw} in order to derive a range of permitted values for $t_{\rm w}$. Making use of the broad range of values derived for GW170817, in Figure~\ref{fig:tw_constrain} we show the allowed (pink)  region  for $t_{\rm w}$ as a function $\dot{M}_{\rm w}$.  The lower limit, which is independent of $\dot{M}_{\rm w}$, is derived from the requirement that $t_{\rm w}/t_{\rm j}\gtrsim 0.7$, as seen in Fig~\ref{fig:ratio}.The upper limit, on the other hand, is derived using equation~\ref{eq:tw} with the additional constraint that $\dot{M}_{\rm w}\gtrsim 10^{-3}M_\odot/{\rm s}$ as  motivated by the range of realistic values seen in merger calculations \citep{Siegel14,2018ApJ...860...64F,2019ApJ...870L..20N, 2014MNRAS.443.3134P}.  A strict upper limit for the time of collapse can be obtained by taking into account the time delay between the GW and the $\gamma$-ray signal, which has been observed to be around $\approx 1.7$s \citep{2017PhRvL.119p1101A,2017ApJ...848L..14G}. The  three constraints can then be combined  to derive a range of permitted values for $t_{\rm w} \approx 1-1.7$s.

This finding  gives further credence to the idea that in the case of GW170817,  the collapse into a black hole was indeed delayed. Yet this argument comes from a completely different line of reasoning than those given in the literature:

\begin{itemize}
\item \citet{2017ApJ...850L..24G} set a constraint of $t_{\rm w} \lesssim 0.9$ s based on the expected lifetime of a hyper-massive NS.
\item  \citet{2018MNRAS.479..588G}  find $t_{\rm w}<1$s based on the time of the shock breakout compared to the observed delay between gravitational waves and the $\gamma$-rays.
\item \citet{2018ApJ...867...18N} perform simulations of a mildly relativistic cocoon using a time delay of $t_{\rm w}\approx0.8-1$s which is able to reproduce the observed data.
\item  \citet{2018MNRAS.479..588G} perform simulations of a cocoon shock breakout and they can reproduce the observed afterglow emission with $t_{\rm w}\approx 1$s.
\item \citet{2018ApJ...856..101M} set a constraint of $t_{\rm w}\approx 0.1-1$s based on the  amount of blue ejecta  expected from a magnetized wind.
\item \citet{2018ApJ...863...58X} perform numerical simulations and show that a delay time of $t_{\rm w}\sim 1$s is able to reproduce the afterglow data
\item \citet{2019ApJ...876..139G} did a comprehensive analysis and estimated the collapse time to be $t_{\rm w}= 0.98^{+0.31}_{-0.26}$s. They combined several constraints including  the delay time between the gravitational wave and electromagnetic signal, a comparison on the observational mass of the blue ejecta and constraints based on a successful jet.
\item \citet{2019ApJ...876L...2V} obtain a limit of $t_{\rm w}\approx 0.67\pm 0.3$s based on observations of extended emission.
\item \citet{2020ApJ...898...59L} favour the delay time to be around $t_{\rm w}<1.1$s by parameter space exploration of jet-wind interactions using an analytical formalism.
\item \citet{2020MNRAS.491.3192H} analytically estimate a delay time of $t_{\rm w}<1.3$s  which they  compared to detailed numerical calculations. 
\end{itemize}

Our estimate for $t_{\rm w}$, which is  based on the angular structure of a successful jet as inferred from afterglow observations, is roughly consistent with these various estimates.

 Many binary neutron star mergers are thought to produce sGRBs when collapsing to black holes  but some merger remnants may experience  significant delays before collapsing.  One expects various outcomes ranging from sGRBs with narrow beams from prompt collapse  to structured jets  with bright and  weak sGRBs for longer collapse timescales.  The properties of the afterglow signatures 
produced by successful and non-successful jets  would provide a natural test to distinguish
between these different progenitor avenues.  The different jet structure can be used to obtain the different afterglow emission \citep{2012ApJ...751...57D,2018MNRAS.478.4553D,2020arXiv201106729U}.

\acknowledgements
We thank  R. Ciolfi, I. Mandel, L. Nativi, O. Gottlieb, R. Margutti, W.-F. Fong, A. Batta, N. Lloyd-Ronning,  C. Kilpatrick, R. Foley, G. Lamb, O. Bromberg,  S.-C. Noble and G. Urrutia for useful discussions  and the referee for useful comments that helped to improve the manuscript.  E.R-R and A.M-B are supported by the  Heising-Simons Foundation, the Danish National Research Foundation (DNRF132) and NSF (AST-1911206 and AST-1852393). A.M-B acknowledges support from a UCMEXUS-CONACYT Doctoral Fellowship and NASA TCAN award TCAN-80NSSC18K1488.
 S.R. acknowledges support by the Swedish Research Council (VR) under grants 2016- 03657\_3 and  2016-06012, the Swedish National Space Board under Dnr. 107/16 and by the Knut and Alice Wallenberg Foundation (KAW 2019.0112). F.D.C. and W.H.L. acknowledge support from the UNAM-PAPIIT grant IG100820. A.J. was supported by the grants no. 2016/23/B/ST9/03114 and 2019/35/B/ST9/04000 from the Polish National
Science Center, and acknowledges computational resources of the Warsaw ICM through grant Gb79-9, and the PL-Grid through the grant grb3. The authors acknowledge use of the lux supercomputer at UC Santa Cruz, funded by NSF MRI grant AST 1828315.

\textit{Software:} yt \citep{2011ApJS..192....9T}.

\bibliography{gw.bib}

\end{document}